\documentstyle[aps,epsfig]{revtex}
\begin{document}
\def\be{\begin{equation}}
\def\ee{\end{equation}}
\def\bfi{\begin{figure}}
\def\efi{\end{figure}}
\def\bea{\begin{eqnarray}}
\def\eea{\end{eqnarray}}

\title{Off-equilibrium generalization of the fluctuation dissipation theorem for Ising spins and measurement of the 
linear response function}

\author{ Eugenio Lippiello$^\dag$, Federico Corberi$^\ddag$
and Marco Zannetti$^\S$}
\address {Istituto Nazionale di Fisica della Materia, Unit\`a
di Salerno and Dipartimento di Fisica ``E.R.Caianiello'',
Universit\`a di Salerno,
84081 Baronissi (Salerno), Italy}
\maketitle

\dag lippiello@sa.infn.it
\ddag corberi@sa.infn.it
\S zannetti@na.infn.it

\begin{abstract}

We derive for Ising spins an off-equilibrium generalization of the fluctuation dissipation
theorem, which is formally identical to the one previously obtained
for soft spins with Langevin dynamics [L.F.Cugliandolo, J.Kurchan and G.Parisi, J.Phys.I France \textbf{4},
1641 (1994)]. The result is quite general and holds both for dynamics with conserved and non conserved order parameter. 
On the basis of this fluctuation dissipation relation, we construct an efficient  
numerical algorithm for the computation of the linear response function without imposing the perturbing field,
which is alternative to those of Chatelain [J.Phys. A \textbf{36}, 10739 (2003)] and Ricci-Tersenghi 
[Phys.Rev.E {\bf 68}, 065104(R) (2003)]. As  applications of
the new algorithm, we present very accurate data for the linear response function of the Ising chain,
with conserved and non conserved order parameter dynamics, finding that in both cases the structure is the
same with a very simple physical interpretation. We also compute the integrated response 
function of the two dimensional Ising model, confirming that it obeys scaling $\chi (t,t_w)\simeq t_w^{-a}f(t/t_w)$,
with $a =0.26\pm 0.01$, as previously found with a different method.

\end{abstract}

PACS: 05.70.Ln, 75.40.Gb, 05.40.-a

\section{Introduction}

In the recent big effort devoted to the understanding
of systems out of equilibrium, of particular interest is the problem of the generalization
of the fluctuation dissipation theorem (FDT).  
The autocorrelation function $C(t-t')$ of some local observable and the
corresponding linear response function $R(t-t')$ in equilibrium are related by the FDT
\be
R(t-t')= {1 \over T}  \frac {\partial C(t-t')}{\partial t'}.
\label{1.1}
\ee
The question is whether an analogous relation 
exists also away from equilibrium, namely whether it is still possible to connect the response
function to properties of the unperturbed dynamics, possibly in the form of correlation functions.

A positive answer to this question exists when the time evolution is of the Langevin type. Consider
a system with an order parameter field $\phi(\vec{x})$ evolving with the equation of motion
\be
{\partial \phi(\vec{x},t) \over \partial t} = B\big (\phi(\vec{x},t) \big ) + \eta (\vec{x},t)
\label{I1}
\ee
where $B\big (\phi(\vec{x},t)\big )$ is the deterministic force and $\eta (\vec{x},t)$ is
a white, zero-mean gaussian noise.
Then, the linear response function is simply given by the correlation function of the
order parameter with the noise
\be
2T R(t,t') = \langle \phi(\vec{x},t) \eta (\vec{x},t')\rangle
\label{I2}
\ee
where $T$ is the temperature of the thermal bath and $t \geq t'$ by causality. It is straightforward~\cite{CKP} to
recast the above relation in the form
\be
TR(t,t')=\frac{1}{2}\frac{\partial C(t,t')}{\partial t'}
-\frac{1}{2}\frac{\partial C(t,t')}{\partial t} -A(t,t')
\label{primi}
\ee
where 
\be
A(t,t')\equiv \frac{1}{2}\left [
\langle \phi(\vec{x},t) B\big (\phi(\vec{x},t')\big ) \rangle- 
\langle B\big (\phi(\vec{x},t)\big )\phi(\vec{x},t')\rangle
\right ]
\label{ax}
\ee
is the so called asymmetry.
Eq.~(\ref{primi}), or~(\ref{I2}),  qualifies as an extension of the FDT out of equilibrium, since in the right hand side
there appear unperturbed correlation functions and, when time translation and time inversion invariance holds, reduces to
the equilibrium FDT~(\ref{1.1}). In Appendix I we show that this equation holds in the same form
both for conserved order parameter (COP) and non conserved order parameter (NCOP) dynamics.

The next interesting question is whether one can do the same also in the case of discrete spin variables, where there is no 
stochastic differential equation and, therefore, Eq.~(\ref{I2}) is not available.
For spin variables governed by a master equation, this problem has been considered 
in recent papers by Chatelain~\cite{chat}, Ricci-Tersenghi~\cite{ricci}, 
Diezemann~\cite{diez} and Crisanti and Ritort~\cite{crisanti}.
However, although important for computational and
analytical calculations, 
their results, as we shall explain below, cannot be regarded as generalizations of the FDT in the sense 
of Eq.~(\ref{primi}). In refs.~\cite{chat,ricci}, a scheme is 
presented where $R(t,t')$ is related to unperturbed correlation functions 
which, however, are not computed in the {\it true} dynamics of the system.
Rather, as will be clarified in Section~\ref{sec5}, these correlation functions are computed through
an auxiliary dynamical rule, instrumental to the construction of an algorithm for the computation
of $R(t,t')$.
On the other hand, in refs.~\cite{diez,crisanti}, $R(t,t')$ is related to
quantities that can be extracted from the
dynamics of the unperturbed system, but not all of them
can be put in the form of correlation functions. 

Instead, we have succeeded in deriving for Ising spin systems a genuine off equilibrium generalization of
the FDT, which takes {\it exactly the same form} as Eqs.~(\ref{primi}) and~(\ref{ax}) and which holds, as
in the Langevin case, for NCOP (spin flip) and COP
(spin exchange) dynamics. Furthermore, using this result we have  
derived an efficient numerical 
method for the computation of the response function, without imposing the
perturbing conjugate field, which is alternative to those of refs.~\cite{chat,ricci}.

The paper is organized as follows: in Section~\ref{sec2} we introduce the formalism
and derive the off equilibrium FDT. 
In Section~\ref{sec3} we introduce the dynamics in
discrete time in order to develop a numerical algorithm based on the fluctuation
dissipation relation.  Section~\ref{sec5} is devoted to a comparative discussion
of our method with those of Chatelain and Ricci-Tersenghi.
In Section~\ref{sec4} the algorithm is applied to the investigation of the scaling properties 
of the response function in the one dimensional Ising model with NCOP (Section~\ref{sec4.1})
and COP dynamics (Section~\ref{sec4.2}). As a further application of the method we study in Sec.~\ref{sec4bis} 
the integrated response function of the Ising model in $d=2$.
In Section~\ref{sec6} we make concluding remarks.

\section{Fluctuation dissipation relation for Ising spins} \label{sec2}

We consider a system of Ising spins $s_i=\pm 1$ executing a markovian stochastic process. The problem is to compute the linear 
response $R_{i,j}(t,t')$ on the spin at the site $i$ and at the time $t$, due to 
an impulse of external field at an earlier time $t'$ and at the site $j$. Let  
\be
h_j(t)=h \delta_{i,j}
\theta (t-t')\theta (t'+\Delta t -t)
\label{pert}
\ee
be the magnetic field on the $j$-th site acting during 
the time interval $[t',t'+\Delta t]$, where $\theta $ is the Heavyside step function. 
The response function then is given by \cite{chat,crisanti}
\be
R_{i,j}(t,t')=\lim_{\Delta t \to 0} \frac{1}{\Delta t}
\left . \frac{\partial \langle s_i(t) \rangle}{\partial h_j(t')} \right
\vert _{h=0}
\label{4}
\ee
where 
\be
\left .\frac{\partial \langle s_i(t) \rangle}{\partial h_j(t')} \right 
\vert _{h=0}=\sum _{[s],[s'],[s'']}
s_i p([s],t\vert [s'],t'+\Delta t) 
\left .\frac{\partial p^h([s'],t'+\Delta t \vert [s''],t')}
{\partial h_j} \right \vert _{h=0} p([s''],t') 
\label{5}
\ee
and $[s]$ are spin configurations.

Let us concentrate on  the factor containing the conditional probability in the presence 
of the external field $p^h([s'],t'+\Delta t \vert [s''],t')$.
In general, the conditional probability for $\Delta t$ sufficiently small is given by
\be
    p([s],t+\Delta t \vert [s'],t)= \delta_{[s],[s']}+
         w([s']\to [s]) \Delta t + {\cal O}(\Delta t^2),
    \label{pippo}
\ee
where we have used the boundary condition $p([s],t\vert [s'],t)= 
\delta_{[s],[s']}$. Normalization of the probability implies 
\be
\sum _{[s']} w([s] \to [s'])=0.
\label{nor}
\ee
Furthermore, the transition rates must verify detailed balance
\be
w([s] \to [s'])\exp(-{\cal H}[s]/T)=w([s'] \to [s])
\exp(-{\cal H}[s']/T),
\label{detbal}
\ee
where ${\cal H}[s]$ is the Hamiltonian of the system. In the following we 
separate explicitly the diagonal from the off-diagonal contributions in 
$w([s] \to [s'])$
\be
 w([s]\to [s'])=-\delta_{[s],[s']}\sum_{[s'']\neq [s]}w([s] \to [s''])
+(1-\delta_{[s],[s']})w([s] \to [s']), 
\label{split}
\ee
where we have used Eq.~(\ref{nor}).

Introducing the perturbing field as an extra term 
$\Delta {\cal H}[s]=-s_j h_j$ 
in the Hamiltonian, to linear order in $h$ the most general form 
of the perturbed transition rates $w^h([s] \to [s'])$ 
compatible with the detailed balance condition is 
(see Appendix II)
\be
w^h([s] \to
[s'])=w^0([s] \to [s'])
\left  \{1-\frac{1}{2 T}h_{j}(s_{j}-s'_j) + M([s],[s'])
\right  \}, 
\label{transh}
\ee
where $M([s],[s'])$ is an arbitrary 
function of order $h/T$ symmetric with respect to the exchange 
$[s] \leftrightarrow [s']$,  
 and $w^0([s] \to [s'])$ are unspecified unperturbed transition rates, which satisfy
detailed balance. 
Notice that, since Eq.~(\ref{detbal}) reduces to an
identity for $[s] \neq [s']$, Eq.~(\ref{transh}) does not hold for 
the diagonal contribution $w^h ([s]\to [s])$ which, in turn, can be obtained by the normalization
condition $\sum _{[s']} w^h([s] \to [s'])=0$.
In the following, for simplicity, we will take $M([s],[s'])=0$ and
the role of a different choice for 
$M([s],[s'])$ will be discussed in sec.\ref{sec5}.

Using Eqs.~(\ref{pippo}),~(\ref{split}) and~(\ref{transh}) we obtain 
\be 
T \left .\frac{\partial p^h([s],t+\Delta t\vert [s'],t)}{\partial h_j}
\right \vert_{h=0} = \Delta t \delta _{[s],[s']}
\frac{1}{2}\sum_{[s'']\neq [s]} w^0([s] \to [s''] )   (s_j-s''_j)
+
\Delta t (1-\delta _{[s],[s']})
\frac{1}{2} w^0([s] \to [s'] )   (s'_j-s_j)
\label{13}
\ee
and inserting this result in Eq.~(\ref{5}), the response function can
be written as the sum of two contributions~\cite{notadded,trap}
\be
    T R_{i,j}(t,t')=\lim _{\Delta t \to 0}\left [T D_{i,j}(t,t',\Delta t)+
T \overline{D}_{i,j}(t,t',\Delta t)\right ],
\label{7}
\ee
where the first term comes from the diagonal part of Eq.~(\ref{13}) 
\be
    T D_{i,j}(t,t',\Delta t)=\frac{1}{2}
     \sum_{[s],[s']}s_ip([s],t\vert
    [s'],t'+\Delta t) \sum_{[s'']\neq [s']}w^0([s']\to [s''])(s'_j-s''_j)
    p([s'],t')
\label{r1r1r1}
\ee
whereas $\overline{D}_{i,j}$ takes all the off-diagonal contributions 
\be
T \overline{D}_{i,j}(t,t',\Delta t)=\frac{1}{2}\sum_{[s],[s'],[s'']\neq[s']}
s_ip([s],t \vert
[s'],t'+\Delta t)(s'_j-s''_j) w^0([s'']\to [s'])p([s''],t') .
\label{erre2}
\ee
Using the time translational invariance of the conditional probability 
$p([s],t\vert[s'],t'+\Delta t)=p([s],t-\Delta t\vert[s'],t')$,
one can write $D_{i,j}(t,t',\Delta t)$ 
in the form of a correlation function
\be
T D_{i,j}(t,t',\Delta t)=-\frac{1}{2} \langle s_i(t-\Delta t)B_j(t')\rangle,
\label{cucu}
\ee
where
\be
B_j=-\sum _{[s'']} (s_j-s''_j) w^0([s]\to [s'']).
\label{bj}
\ee
Using Eqs.~(\ref{pippo}) and~(\ref{split}) the off-diagonal contribution can be written as
\be
T \overline{D}_{i,j}(t,t',\Delta t)=
\frac{1}{2}\frac{\Delta C_{i,j}(t,t')}{\Delta t}
\label{deltac}
\ee 
where
\begin{eqnarray}
& & \Delta C_{i,j}(t,t')=\langle s_i(t)[s_j(t'+\Delta t)-s_j(t')]\rangle=
\label{dcc} \nonumber \\
& &  \sum _{[s],[s'],[s'']}s_i(s_j'-s_j'')p([s],t\vert[s'],t'+\Delta t)
p([s'],t'+\Delta t\vert[s''],t')p([s''],t').
\label{re}
\end{eqnarray}
Therefore, putting together Eqs.~(\ref{cucu}),(\ref{deltac}) and
taking the limit $\Delta t \to 0$ we obtain
\be
T R_{i,j}(t,t')=\frac{1}{2}\frac {\partial C_{i,j}(t,t')}{\partial t'}-
\frac{1}{2}\langle s_i(t)B_j(t')\rangle.
\label{nuova}
\ee

In order to bring this into the form of Eqs.~(\ref{primi}) and~(\ref{ax}), 
we notice that from Eqs.(\ref{pippo},\ref{split}) follows  
\begin{eqnarray}
& & \frac{d\langle s_j(t)\rangle}{dt}=\sum _{[s]}s_j \frac{dp([s],t)}{dt} 
\label{labbbbb} \nonumber \\
& & =-\sum _{[s]}\sum_{[s'']\neq [s]}
s_jw^0([s]\to[s''])p([s],t)
+\sum _{[s],[s']\neq [s]}
s_j w^0([s']\to [s]) p([s'],t) .
\label{labe}
\eea
Hence, after the change of variables $[s] \to [s''], [s'] \to [s]$ in the second sum,
one obtains

\be
\frac{d\langle s_j(t)\rangle}{dt}
=-\sum _{[s]}\sum _{[s'']\neq [s]}
(s_j-s''_j) w^0([s]\to [s'']) p([s],t) = \langle B_j(t) \rangle.
\label{ala}
\ee
In a similar way, it is straightforward to derive
\be
\frac {\partial C_{i,j}(t,t')}{\partial t} -
\langle B_i(t)s_j(t') \rangle=0
\label{nuova2}
\ee
and subtracting this from Eq.~(\ref{nuova}) we finally find
\be
T R_{i,j}(t,t')=\frac{1}{2}\frac {\partial C_{i,j}(t,t')}{\partial t'}-
\frac{1}{2}\frac {\partial C_{i,j}(t,t')}{\partial t}-A_{i,j}(t,t')
\label{treb}
\ee
where $A_{i,j}(t,t')$ is given by
\be
A_{i,j}(t,t')=\frac{1}{2}\left [\langle s_i(t)B_j(t')\rangle-\langle B_i(t)s_j(t') \rangle \right ].
\label{asym}
\ee
Eqs.~(\ref{treb}) and~(\ref{asym}) are the main result of this paper. They are 
identical to Eqs.~(\ref{primi}) and~(\ref{ax}) for Langevin dynamics, since the observable $B$ entering in the
asymmetries~(\ref{ax}) and~(\ref{asym}) plays the same role in the two cases. In fact, Eq.~(\ref{ala})
is the analogous of 
\be
{\partial \langle \phi(\vec{x},t)\rangle \over \partial t} = \langle B\big (\phi(\vec{x},t)\big )\rangle
\label{bb4}
\ee
obtained from Eq.~(\ref{I1}) after averaging over the noise.

In summary, Eq.~(\ref{treb}) is a relation between the 
response function and correlation functions of the unperturbed
kinetics, which generalizes the FDT. 
Furthermore, Eq.~(\ref{treb}) applies to a wide class of systems.
Besides being obeyed by soft and hard spins,
it holds both for COP and NCOP
dynamics. Moreover, as it is clear by its derivation,
Eq.~(\ref{treb}) does not require any 
particular assumption 
on the hamiltonian nor on the form of the unperturbed transition rates.

\section{Dynamics in discrete time: the numerical algorithm}\label{sec3}

We now discuss the numerical implementation of
the fluctuation dissipation relation derived above, as a method to compute the response function
without imposing the external magnetic field~(\ref{pert}).
Let us recall that Eq.~(\ref{nuova}) was obtained letting $\Delta t \rightarrow 0$
\be
T R_{i,j}(t,t') = \frac{1}{2}\lim _{\Delta t\to 0}\left [
\frac{\Delta C_{i,j}(t,t')}{\Delta t}- 
\langle s_i (t-\Delta t) B_j(t') \rangle \right ]. 
\label{3.919} 
\ee 
In the simulations 
of an $N$-spin system, time is discretized by the elementary
spin updates. Measuring time in montecarlo steps,
the smallest available time $\epsilon =1/N$ 
is the one associated to a single update. 
Then, in discrete time Eq.~(\ref{3.919}) reads
\be
T R_{i,j}(t,t') = \frac{1}{2}\frac {C_{i,j}(t,t'+\epsilon) - C_{i,j}(t,t')}{\epsilon} -
\frac {1}{2}\langle s_i (t-\epsilon) B_j(t') \rangle 
\label{3.91} 
\ee 
and we use this for the numerical calculation of the response function.
For completeness we also give the expression for the
integrated response function     
\be
\chi_{i,j}(t,[\overline t,t_w])=\int _{t_w}^{\overline t} R_{i,j}(t,t')dt'
\ee
which correspond to the application of
a constant field between the times $t_w$ and $\overline t$.
From Eq.~(\ref{3.91}) we have
\be
T \chi_{i,j}(t,[\overline t,t_w]) =  T \epsilon \sum_{t'=t_w}^{\overline t}
R_{i,j}(t,t') = \frac{1}{2}[C_{i,j}(t,\overline t) - C_{i,j}(t,t_w)] -
\frac{\epsilon}{2}\sum_{t'=t_w}^{\overline t}\langle s_i(t-\epsilon) B_j(t') 
\rangle 
\label{new}
\ee
where $t-\epsilon \geq \overline t > t_w \geq 0$,
and $ \sum_{t'=t_w}^{\overline t}$ stands for the sum over the discrete
times in the interval $[t_w,\overline t]$.

\section{Comparison with different algorithms}\label{sec5}

Expressions for the response
function in a discretized time dynamics have been derived previously
by Chatelain~\cite{chat} and Ricci-Tersenghi~\cite{ricci}. 
Restricting to transition rates of the heat-bath form and to 
the case of single flip dynamics they have obtained
\begin{eqnarray}
T \chi_{i,j}(t,[\overline t,t_w]) & = & \sum _{I(\overline t,t_w)}
\sum_{\tau =t_w}^{\overline t} \delta _{j,I(\tau)}\langle s_i(t)(s_j(\tau)-s^W_j(\tau))\rangle 
_{I(\overline t,t_w)},
\label{old}
\end{eqnarray}
where $I(\tau)$ is the index of the site updated at the discrete time $\tau $,
$I(\overline t,t_w)$ is a specific sequence of $I(\tau)$'s between the
times $t_w$ and $\overline t$, $s_j^W(\tau) =
\tanh [\beta h_j^W(\tau )]$ and $h_j^W(\tau)$ is the local field
due to the spins interacting with $s_j$.

It is important to stress the differences between Eqs.~(\ref{old}) 
and~(\ref{new}). Although in the r.h.s. of Eq.~(\ref{old}) there appears
an unperturbed correlation function,
this is computed in the {\it ad hoc} kinetic rule introduced
for the purpose of evaluating the response function. 
In Eq.~(\ref{new}), instead, the average in the r.h.s. is computed in
the true unperturbed dynamics of the system.
This important difference arises because of
the presence of the delta function $\delta _{j,I(\tau)}$ in
Eq.~(\ref{old}), which constrains to update at the time $\tau$ only the spin at the site $j$,  
where the external field is applied.
Then, in the averaging procedure, only the subset of dynamical trajectories 
with $I(\tau)=j$ are considered, while all the others get zero
statistical weight, which is not what happens in the true unperturbed dynamics. 
Eq.~(\ref{old}), therefore, although useful for the computation
of the response function, is operatively
restricted to a numerical protocol with 
a sequential updating satisfying the constraint imposed by
the delta function $\delta _{j,I(\tau)}$. The correlations functions
appearing in this equation cannot be extracted numerically from the
behavior of the original unperturbed system, and cannot be accessed in an experiment.  

Another difference between Eq.~(\ref{old}) and Eq.~(\ref{new}) concerns the choice of
$M([s],[s'])$. Our results are obtained with $M([s],[s'])=0$. 
Instead, Eq.~(\ref{old}) corresponds to $M([s],[s'])\ne 0$. 
In fact, Eq.~(\ref{old}) assumes heat bath transition rates
$w([s]\to[s'])=\{\exp [-{\cal H}[s']/T]\}/\{\exp [-{\cal H}[s]/T] +\exp [-{\cal H}[s']/T]\}$;
expanding this expression to first order in powers of $h/T$, and comparing with Eq.~(\ref{transh}),
one has 
\be
M([s],[s'])=\frac{h_j}{2T}(s'_j-s_j)
\frac{e^{{\cal H}[s']/T}-e^{{\cal H}[s]/T}}{e^{{\cal H}[s']/T}+e^{{\cal H}[s]/T}}.
\ee

Retaining $M([s],[s'])$ in Eq.~(\ref{transh}) and following the same steps as
for $M([s],[s'])=0$, one finds the extra term
\be
\Delta R(t,t')=\frac{T}{h_j}\sum _{[s],[s'],[s'']}s_ip([s],t\vert [s'],t')M([s'],[s''])
\left [ w^0([s'']\to[s'])p([s''],t')-w^0([s']\to[s''])p([s'],t')
\right ]
\ee
in addition to the quantities already present on the r.h.s. of Eq.~(\ref{treb}).
This term cannot be related to
correlation functions.
It is generally believed that the
large scale - long time properties of the dynamics
(perturbed or not) do not depend too much, within a given universality class, 
on the form of the transition rates. 
Then one expects the corrections $\Delta R(t,t')$
introduced by different choices of 
$M([s],[s'])$ to be negligible. 
Indeed, as will be shown in the following sections~\ref{sec4},\ref{sec4bis},
numerical results obtained for the Ising model in $d=1,2$ 
with the two algorithms are not sensitive to the choice of  
$M([s],[s'])$ . 


\section{Response function of the Ising chain}\label{sec4}

As an application of the numerical method, we 
compute the autoresponse function $R(t,t')=R_{i,i}(t,t')$ in the $d=1$
Ising model, with and without conservation of the order parameter.  
We consider the system prepared in the infinite temperature equilibrium state 
and quenched to the finite temperature $T>0$ at the time $t=0$. 
Since the critical temperature vanishes for $d=1$, the final correlation length $\xi_{eq}$
is finite and equilibrium is reached in a finite time $\tau _{eq}\sim \xi_{eq}^z$, 
where $z$ is the dynamic exponent. For deep quenches $\tau _{eq}$ is large and, 
after a characteristic time $t_{sc}$, 
a well defined non-equilibrium scaling regime sets in for 
$t_{sc}<t<\tau _{eq}$, characterized by the growth of the domains
with a typical size $L(t)\sim t^{1/z}$. We study the scaling
properties of the response function $R(t,t')$ when
both $t$ and $t'$ belong to the scaling regime.

\subsection{Non conserved dynamics}\label{sec4.1}

The linear response function in the model with single spin flip dynamics has been computed
analytically~\cite{noi1,hen,mayer}. This case, therefore,
is useful as a test for the accuracy of the algorithm.
In the aging regime $t' \leq t \leq \tau_{eq}$  the autoresponse function $R(t,t')$ is given 
by~\cite{hen} 
\be
T R(t,t') =
e^{-(t-t')}I_0\left ( t-t'\right )e^{-2t'}\left [
I_0\left (2t'\right) +I_1\left (2t'\right ) \right ]   
\label{d1.1} 
\ee 
where $I_n(x)$  are the modified Bessel functions. 
  
In order to improve the signal to noise ratio, we have extracted $R(t,t')$ 
from the integrated autoresponse function 
$\chi(t,[t'+\delta,t'])$ by choosing $\delta$ in the following way.
Expanding  for small $\delta$ we have
\be
\frac{\chi(t,[t'+\delta,t'])}{\delta}\simeq 
R(t,t')+ \frac{\delta}{2}\frac{\partial R(t,t')}{\partial t'}
\label{ap}
\ee
then, for a given level of accuracy, $R(t,t')$ can be
obtained from $\chi(t,[t'+\delta,t'])/\delta$
if $\delta$ is chosen appropriately small. Notice that,
assuming scaling $R(t,t')=t^{-(a+1)}f(t'/t)$, from Eq.~(\ref{ap})
one has that, for a given value of $x=t'/t$, the second term on 
the r.h.s. produces a relative correction 
$\Delta R(t,t')/R(t,t')=(1/2)[f'(x)/f(x)](\delta/t)$
of order $\delta/t$.
In our simulations we have chosen $\delta=1$ and, since the simulated times
are $t\geq 100$, we have always $\delta/t \leq 10^{-2}$. In the
following it is understood that all numerical results for $R(t,t')$ are 
obtained in this way.

In Fig.~{\ref{rncop} we compare the numerical results obtained by means of
the algorithm~(\ref{new}), 
for three different values of $t'$,
with the exact solution~(\ref{d1.1}). We have also plotted the data
obtained using the algorithm~(\ref{old}) of Ricci-Tersenghi~\cite{ricci}, finding
an excellent agreement between the 
curves generated by the different algorithms and the analytical
expression~(\ref{d1.1}). 

The physical meaning of the exact solution can be understood by replacing Eq.~(\ref{d1.1}) with the simple
interpolating formula  
\be
TR(t,t') = A_z t'^{-1/z}(t-t'+t_0)^{1/z-1}
\label{rapprox}
\ee
obtained by replacing the Bessel functions with the dominant term in the asymptotic expansion and
by inserting $t_0$ as a regularization of $R(t,t')$ at equal times. For NCOP
$z=2$. With $A_2=1/(\sqrt 2 \pi)$ and $t_0=1/(2\pi)$
the simple algebraic form~(\ref{rapprox}) gives a very good approximation of the exact solution all over the
time domain, from short to large time separations. 
Rewriting  Eq.~(\ref{rapprox}) as
\be
TR(t,t') \sim L(t')^{-1} TR_{sing}(t-t') 
\label{d1.3} 
\ee 
where
\be
TR_{sing}(t-t')=A_z (t-t')^{1/z-1}
\label{dd1.3}
\ee
the physical meaning becomes clear, since $L(t')^{-1}$ 
is proportional to the density of
defects  $\rho (t')$ at time $t'$, and $R_{sing}(t-t')$
can be interpreted as the response associated to a single defect.
In other words, Eq.~(\ref{d1.3}) means that the
total response is given by the contribution of a single defect times
the density of defects at the time $t'$.
As a matter of fact, Eqs.~(\ref{d1.3}) and~(\ref{dd1.3}) are the particular realization 
of a general pattern for the aging part of the response function~\cite{Corberi2003}
\be
R(t,t') \sim L(t')^{-1} R_{sing}(t-t')f(t/t').
\label{rgeneral}
\ee
The presence of the scaling function $f(t/t')$ in Eq.~(\ref{rgeneral})
for $d>1$ can be explained as follows:
in $d=1$ interfaces are point like and the interaction between
them always produces annihilation. This is accounted for
by the defect density  $L(t')^{-1}$, so $f(t/t')\equiv 1$. 
In higher dimension, however,
defects are extended objects whose interaction can produce
a wealth of different situations, which are globally described
by a suitable scaling function $f(t/t')$.

\subsection{Conserved dynamics}\label{sec4.2}

The generality of the structure of Eqs.~(\ref{d1.3}) and~(\ref{dd1.3}) 
may soon be tested by looking at the response function in the Ising chain with 
COP dynamics.
While for NCOP the system enters the scaling regime almost immediately,
since $t_{sc}\simeq 1$, for COP  
the time $t_{sc}$ for the onset of the
scaling regime is of the order of the characteristic time $\tau_{ev} \sim \exp(4 J/T)$ for the separation 
(evaporation) of a
spin from the boundary of a domain.
The equilibration time, instead, is given by
$\tau _{eq}\sim \exp(10 J/T)$~\cite{cornell}.
In order to have a large scaling regime, namely $\tau_{eq} \gg
t_{sc}$, it is necessary to take $T/J \ll 1$, and to choose $t_w>t_{sc}$. 
Simulations of the system in these conditions are excessively
time demanding with a conventional montecarlo algorithm. 
Therefore we have resorted to the fast algorithm of 
Bortz, Kalos and Lebowitz
\cite{bkl}, which is much more efficient at low temperatures.
With a conventional algorithm a number of attempts
proportional to $\tau_{ev}$ is necessary on average before the
evaporation of a spin from a domain occurs. Then, at low
temperatures, a huge amount of attempted moves are rejected,
causing a very low efficiency. The algorithm of Bortz, Kalos
and Lebowitz, instead, is rejection free: moves are
always accepted and time is increased proportionally to the
inverse probability associated with them.
We stress that this is not an approximate kinetics, but a
clever implementation of the exact dynamics. 

From the
unperturbed system the response function is extracted 
through Eq.~(\ref{new}), as in the case of NCOP. About the choice of $\delta$, for COP the simulated times are $t\geq 10^7$
and we can have $\delta/t\leq 10^{-2}$, as required for the
correction term in Eq.~(\ref{ap}) to be negligible, with $\delta =10^5$.
 We have performed simulations with $T=0.3J$, corresponding to
$t_{sc} \simeq 6\cdot 10^5$ and $\tau _{eq}\simeq 3\cdot 10^{14}$.
In this conditions the scaling regime is very large.
Actually, after
a very narrow initial regime,  
where single spins diffuse until
they are adsorbed on an interface, no evaporations occur and 
the system is frozen up to times of order $\tau _{ev}$. 
In this time regime the density of defects $\rho (t)$
stays constant, as shown in Fig.~\ref{rho}. Then, for $t\gtrsim \tau _{ev}$,
the evaporation-condensation mechanism takes place and the systems
gradually enters the scaling regime. The range of times explored for
the computation of the response function is shown in 
Fig.~\ref{rho}. This has been chosen as a compromise between
the necessity to go to the largest accessible times, in order to
work well inside the scaling regime, and to speed up the simulation
to have a good statistics. For COP, the observation of the asymptotic behavior $\rho (t)\sim t^{1/z}$,
with $z=3$, requires very large time~\cite{effexp}. In the range of times
explored for the computation of $R(t,t')$ the effective exponent
$z_{eff}=-[d\log \rho (t)/d\log t]^{-1}$ has reached the value $z_{eff}= 3.44$.  

We have plotted in Fig.~\ref{rcop} the numerical data for $R (t,t')$
together with the curves obtained from the analytical form~(\ref{rapprox}),
where we have substituted for $z$ the above value of $z_{eff}$ and we have used
$A_3$ and $t_0$ as fitting parameters. The comparison is good
and suggests that the physical interpretation, behind
the form~(\ref{d1.3}) and~(\ref{dd1.3}) of the 
response function, applies also to the $d=1$ Ising
model with spin exchange dynamics~\cite{nota}.
We expect the more general form~(\ref{rgeneral})
to hold in higher dimension with COP. 

\section{Response function of the $d=2$ Ising model}\label{sec4bis}

As a further application of the numerical method, we 
compute the zero field cooled magnetization (ZFC) $\chi(t,t_w)=\chi(t,[t,t_w])$ in the $d=2$
Ising model with NCOP, quenched from the infinite temperature equilibrium state 
to a temperature below $T_c$. This quantity has already been
measured both by applying the perturbation~\cite{Corberi2003,barrat,noi2,noi3} or by means of the
algorithm of Ricci-Tersenghi~\cite{ricci}. In Fig.~\ref{figchi1} we compare results
obtained with our method and with that of Ricci-Tersenghi, for several values of
$t_w$ in the scaling regime. The agreement between the two algorithms is excellent also
in this case. The equivalence of the two algorithms both in $d=1$ and in $d=2$ suggests,
recalling the discussion at the end of Sec.~\ref{sec5}, that
different choices of $M([s],[s'])$ do not produce significant differences.

Let us comment on the behavior of $\chi(t,t_w)$.
As it is well known, in the late stage of phase-ordering the interior of the
growing domains is equilibrated, while interfaces are out of equilibrium.
Then, a distinction can be made between bulk and interface fluctuations.
Accordingly, for the ZFC one has~\cite{bouchaud,Franz,ninfty}
\be
\chi(t,t_w)=\chi_{st}(t,t_w)+\chi_{ag}(t,t_w).
\label{chichi}
\ee
Here $\chi_{st}(t,t_w)$ is the contribution from the bulk of domains, which
behaves as the equilibrium response $\chi_{eq}(t,t_w)$ in the ordered state at the
temperature $T$. This quantity, starting from zero at $t=t_w$, saturates to
the value $1-M^2$, $M$ being the equilibrium magnetization. 
The other term appearing in Eq.~(\ref{chichi}), namely the additional
aging contribution due to the interfaces, is much less known. It is expected to scale as
\be
\chi_{ag}(t,t_w)=t_w^{-a}f(\frac{t}{t_w}).
\label{scscal}
\ee

In previous studies~\cite{Corberi2003,noi2,noi3},
an auxiliary dynamics, which prevents flips in the bulk,
was used in order to extract the aging part of the response in Eq.~(\ref{chichi}).
Here, instead, we have chosen a different technique to isolate $\chi_{ag}(t,t_w)$:
We compute the full $\chi (t,t_w)$ in the
Glauber dynamics working at a sufficiently low temperature where
$\chi_{st}(t,t_w)$ is negligible. In fact, by choosing $T=J$, the asymptotic
value of $\chi_{st}(t,t_w)$ is $1-M^2\simeq 0.0014$, much smaller than
the computed values of $\chi(t,t_w)$ in the range of times considered. Then one has
$\chi (t,t_w)\simeq \chi_{ag}(t,t_w)$.

Besides this difference, previous results~\cite{Corberi2003,noi2,noi3} 
on $\chi _{ag}(t,t_w)$ were obtained with the usual method where a perturbation is applied.
The strength $h$ of the perturbation must be chosen sufficiently small to work in the
linear regime. However, by reducing $h$ the signal to noise ratio lowers, and
the results get worst.
Then one usually runs a series of preliminary simulations in order to determine the
largest value of $h$ compatible with the requisite of working in the linear regime. 
While this point may be subtle, in the result presented in this paper
the limit $h\to 0$ is taken analytically in the derivation of the algorithm.

We have extracted $a$ from the data of Fig.~\ref{figchi1} by plotting
$\chi (t,t_w)$ against $t_w$ with $x=t/t_w$ held fixed. The results are shown 
on a double logarithmic plot in Fig.~\ref{figchi2}. According to the scaling form~(\ref{scscal}), 
for different values of $x$ the data must align on straight lines
with the same slope $a$. This is very well compatible with the curves of
Fig.~\ref{figchi2}, indicating that scaling is obeyed. Computing
$a$ as the slope of these curves we find $a=0.26 \pm 0.01$.
This result agree very well with the value found in~\cite{Corberi2003,noi2,noi3}.
Once this exponent is known, one obtains data collapse by plotting
$t_w^{a}\chi (t,t_w)$ against $x$, as shown in the inset of Fig.~\ref{figchi2},
confirming the validity of the scaling form~(\ref{scscal}).

\section{Concluding remarks}\label{sec6}

In this paper we have derived a generalization of the FDT out of equilibrium for systems of
Ising spins, which takes exactly
the same form of the FDT generalization previously
derived~\cite{CKP} for soft spins, evolving with Langevin dynamics.

We have shown that this fluctuation dissipation relation, which reduces to
the usual FDT when equilibrium is reached, 
is obeyed in complete generality both by systems with COP and
NCOP. 
In addition to the theoretical interest, as a contribution to the understanding of
the FDT in the out of equilibrium regime,
our result is promising also as a convenient tool for the computation of the linear
response function in numerical simulations without applying the
perturbation, along the line of refs.~\cite{chat,ricci}.
With standard methods, the requirement to work
in the linear regime, namely with an adequately small perturbation,  
sometimes is very subtle and hard to be checked. 
This problem is avoided by this new class of algorithms.
Moreover, the statistical
accuracy of the results is, for comparable cpu times, much better
because simulations of perturbed systems usually require
additional statistical averages over realizations of the (random) perturbation. 
We have demonstrated the high quality
of the results produced by our algorithm by computing the response function
of the Ising model in $d=1$ and the integrated response function in $d=2$. 
In $d=2$ our results agree with those obtained with the
algorithm of ref.~\cite{ricci} and with previous simulations performed 
applying the perturbation. We confirm that $\chi (t,t_w)$ obeys a scaling
form~(\ref{scscal}) with $a =0.26 \pm 0.01$, in agreement with previous 
determinations~\cite{Corberi2003,noi2,noi3}
of this exponent.
In $d=1$, for NCOP our
results are in excellent agreement with the exact analytical
solution and with the simulation made with the algorithm 
of ref.~\cite{ricci}. In the case of COP,
where no analytical solution is available,
we have obtained results which substantiate the existence of the common
structure~(\ref{d1.3},\ref{dd1.3}) of the response function for COP and NCOP. 
These results show that the algorithm is 
efficient enough to give access to the direct measurement of the impulsive response function $R(t,t')$,
which is too noisy to be computed with standard methods.
For this reason, previous numerical studies~\cite{Corberi2003,barrat,Integrated,berthier} 
have been necessarily directed to the
investigation of the integrated
response functions, such as the thermoremanent magnetization or 
the zero field cooled magnetization, which are easier to  compute. 
However, as discussed in detail in~\cite{Corberi2003}, it is quite
delicate a task to extract
the properties of $R(t,t')$ 
from those of the integrated response functions.  
Therefore, the feasibility of a direct computations of $R(t,t')$ is an important development in the
field, which is expected to solve a number of problems still open~\cite{Corberi2003} on the scaling behavior
of $R(t,t')$ for $d>1$.

\vspace{1cm}
\centerline{\bf ACKNOWLEDGMENTS}
\vspace{.6cm}
We are much indebted to Claudio Castellano for valuable suggestions
on the numerical techniques. 
This work has been partially supported
from MURST through PRIN-2002.

\vspace{1cm}
{\noindent \bf APPENDIX I}
\vspace{.6cm}

Let us write the Langevin equation in the general form
\be
{\partial \phi(\vec{x},t) \over \partial t} = (i\nabla)^p\left [  B\big (\phi(\vec{x},t)\big ) 
+ h(\vec{x},t) \right ] + \eta (\vec{x},t)
\label{APP1}
\ee
where  $h(\vec{x},t)$ is the external field conjugated to the order parameter, $p=0$ or $p=2$ for NCOP or COP, 
respectively, and the noise correlator is given by
\be
\langle \eta (\vec{x},t)\eta (\vec{x}',t')\rangle = (i\nabla)^p 2T \delta(\vec{x}- \vec{x}')\delta(t-t').
\label{APP2}
\ee
Fourier transforming with respect to space, these become
\be
{\partial \phi(\vec{k},t) \over \partial t} = k^p  \left [ {\cal B}\left ([\phi],\vec{k},t\right ) 
+ h(\vec{k},t) \right ] + \eta (\vec{k},t)
\label{APP3}
\ee
and
\be
\langle \eta (\vec{k},t)\eta (\vec{k}',t')\rangle =  2T k^p (2\pi)^d\delta(\vec{k}+ \vec{k}')\delta(t-t')
\label{APP4}
\ee
where  ${\cal B}\left ([\phi],\vec{k},t\right )$ is the Fourier transform of $B\left (\phi(\vec{x},t)\right )$.

The linear response function is defined by
\be
R(\vec{k},t,\vec{k}',t') = \left . {\delta \langle \phi(\vec{k},t) \rangle_h \over \delta h(\vec{k}',t')}
\right |_{h=0}
\label{APP5}
\ee
with $t \geq t'$.
Notice that, since $\eta (\vec{k},t)$ and  $k^p h(\vec{k},t)$ enter the equation of motion~(\ref{APP3})
in the same way, we have
\be
\left \langle {\delta \phi(\vec{k},t) \over \delta \eta(\vec{k}',t')} \right \rangle =
{ 1 \over k'^p} \left . {\delta \langle \phi(\vec{k},t) \rangle_h \over \delta h(\vec{k}',t')}
\right |_{h=0}
\label{APP6}
\ee
where $\langle \cdot \rangle$ denotes averages in absence of the external field. Then,
using the identity~\cite{Zinn}
\be
\langle  \phi(\vec{k},t)\eta(\vec{k}',t') \rangle = 2Tk'^p 
\left \langle {\delta \phi(\vec{k},t) \over \delta \eta(\vec{k}',t')} \right \rangle
\label{APP7}
\ee
we find
\be
2TR(\vec{k},t,\vec{k}',t') = \langle  \phi(\vec{k},t)\eta(\vec{k}',t') \rangle 
\label{APP8}
\ee
or in real space
\be
2TR(\vec{x},t,\vec{x}',t') = \langle  \phi(\vec{x},t)\eta(\vec{x}',t') \rangle 
\label{APP9}
\ee
showing that Eq.~(\ref{I2}) holds in the same form for NCOP and COP.

\vspace{1cm}
{\noindent \bf APPENDIX II}
\vspace{.6cm}

The transition rates, with and without the external field, satisfy the detailed balance condition~(\ref{detbal}). 
Writing $w^h([s]\to [s'])= w^0([s]\to [s'])\,\Delta w([s]\to [s'])$ and
${\cal H}^h[s]={\cal H}^0[s]+\Delta {\cal H}[s]$, from Eq.~(\ref{detbal}) 
follows
\be
\Delta w([s]\to [s'])\exp \left [-\frac {\Delta {\cal H}[s]}{T} \right ]=
\Delta w([s']\to [s])\exp \left [-\frac{\Delta {\cal H}[s']}{T}\right ].
\label{detta}
\ee
Using $\Delta {\cal H}[s]=-s_j h_j$, Eq.~(\ref{detta})
is satisfied by
$\Delta w([s]\to [s'])=\exp \left [-(1/(2T))h_{j}(s_{j}-s'_j) \right ]$
up to a factor ${\cal M}([s],[s'])$ which satisfies ${\cal M}([s],[s'])={\cal M}([s'],[s])$.
Therefore, the most general form of the perturbed
transition rates, compatible with detailed balance, is given by
\be
\Delta w([s]\to [s'])=\exp \left [-\frac{1}{2T}h_{j}(s_{j}-s'_j)\right ]
\,{\cal M}([s],[s']).
\label{detta2}
\ee
For $h=0$ the condition $\Delta w([s]\to [s'])=1$ implies
${\cal M}([s],[s'])=1$.
Therefore, to linear order in the perturbation one has
${\cal M}([s],[s'])\simeq
1+M([s],[s'])$. Inserting this result in Eq.~(\ref{detta2}), and expanding 
also the exponential term, to linear order in $h/T$ one obtains Eq.~(\ref{transh}).

\newpage

\begin{figure}
    \centering
   \rotatebox{0}{\resizebox{.65\textwidth}{!}{\includegraphics{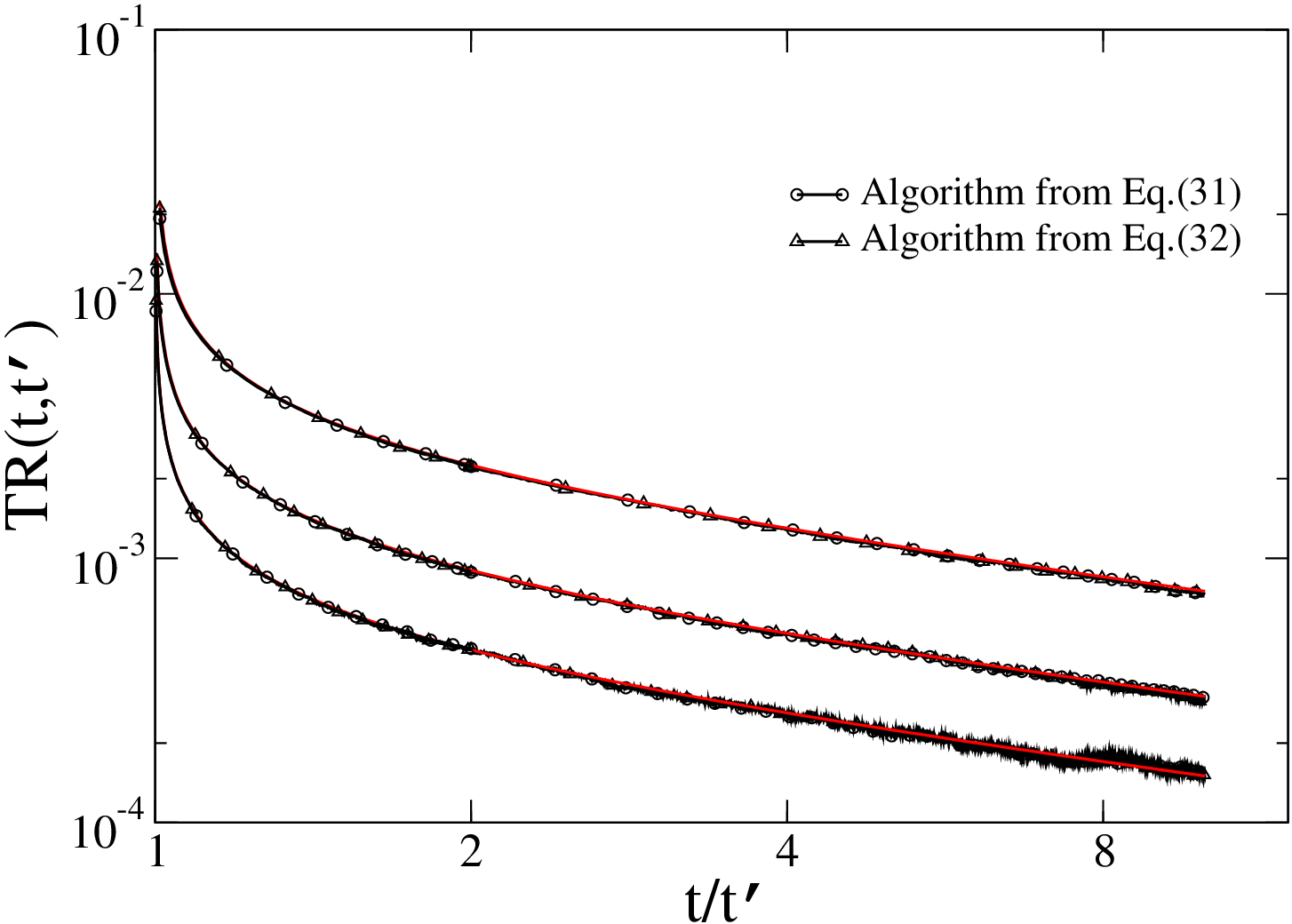}}}
    \caption{$R(t,t')$ in the $d=1$ Ising
    model with NCOP, $T=0.3 J$ and $J=1$. The number of spins is $N=10^4$,
    $t'=100, 250, 500$ (mcs) from top to bottom.
    Data from different algorithms correspond to
    different symbols. 
    Continuous curves are the plots of Eq.~(\ref{d1.1})}.

\label{rncop}
\end{figure}

\newpage

\begin{figure}
    \centering
   \rotatebox{0}{\resizebox{.65\textwidth}{!}{\includegraphics{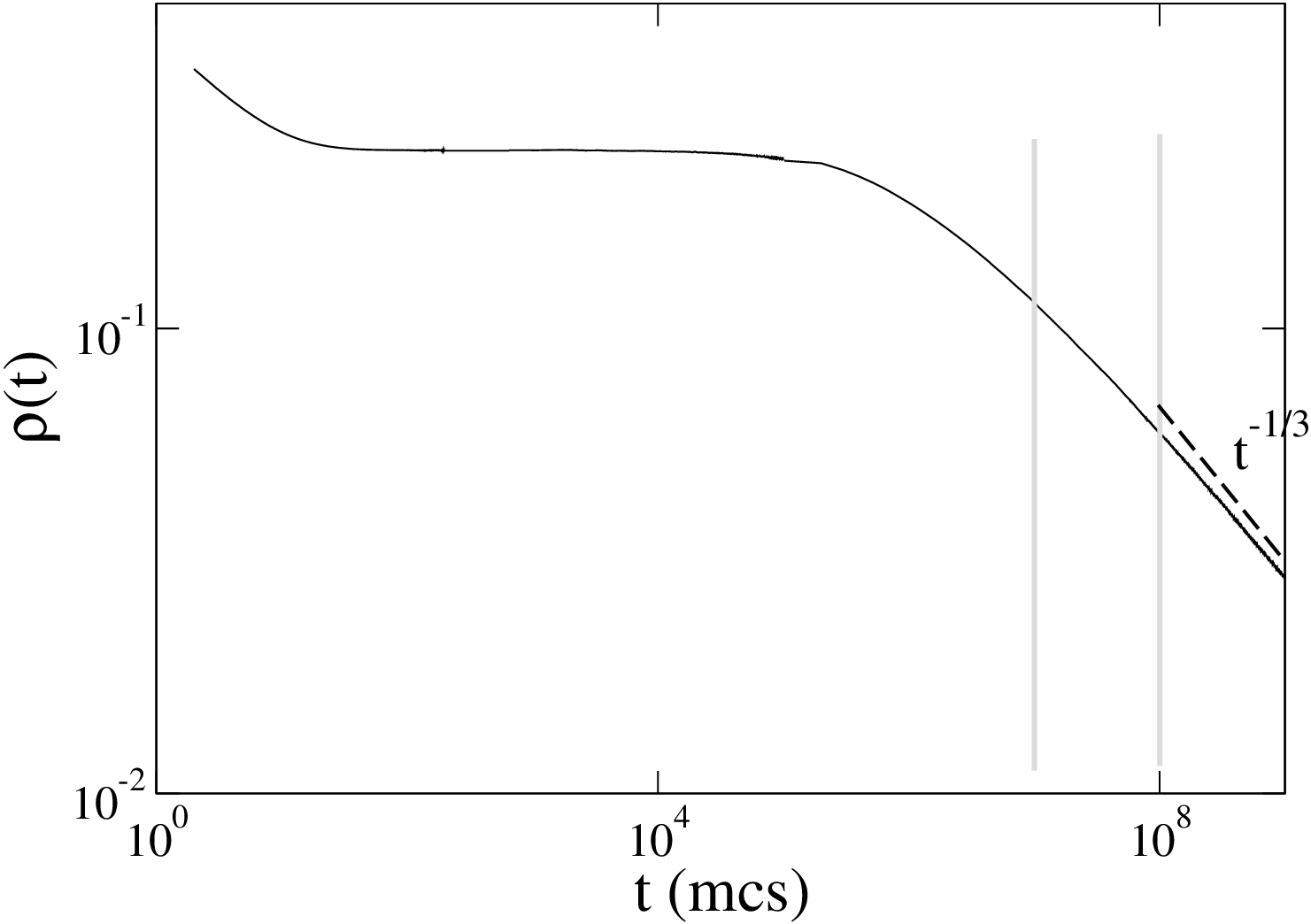}}}
    \caption{$\rho (t)$ in the $d=1$ Ising
    model with COP, $T=0.3 J$ and $J=1$. The number of spins is $N=10^4$. The range of time 
    used in the simulations for the computation
    of the response function is in between the vertical lines. The dashed line represents the
    asymptotic law $\rho (t)\sim t^{-1/3}$.}
\label{rho}
\end{figure}

\newpage

\begin{figure}
    \centering
    \rotatebox{0}{\resizebox{.65\textwidth}{!}{\includegraphics{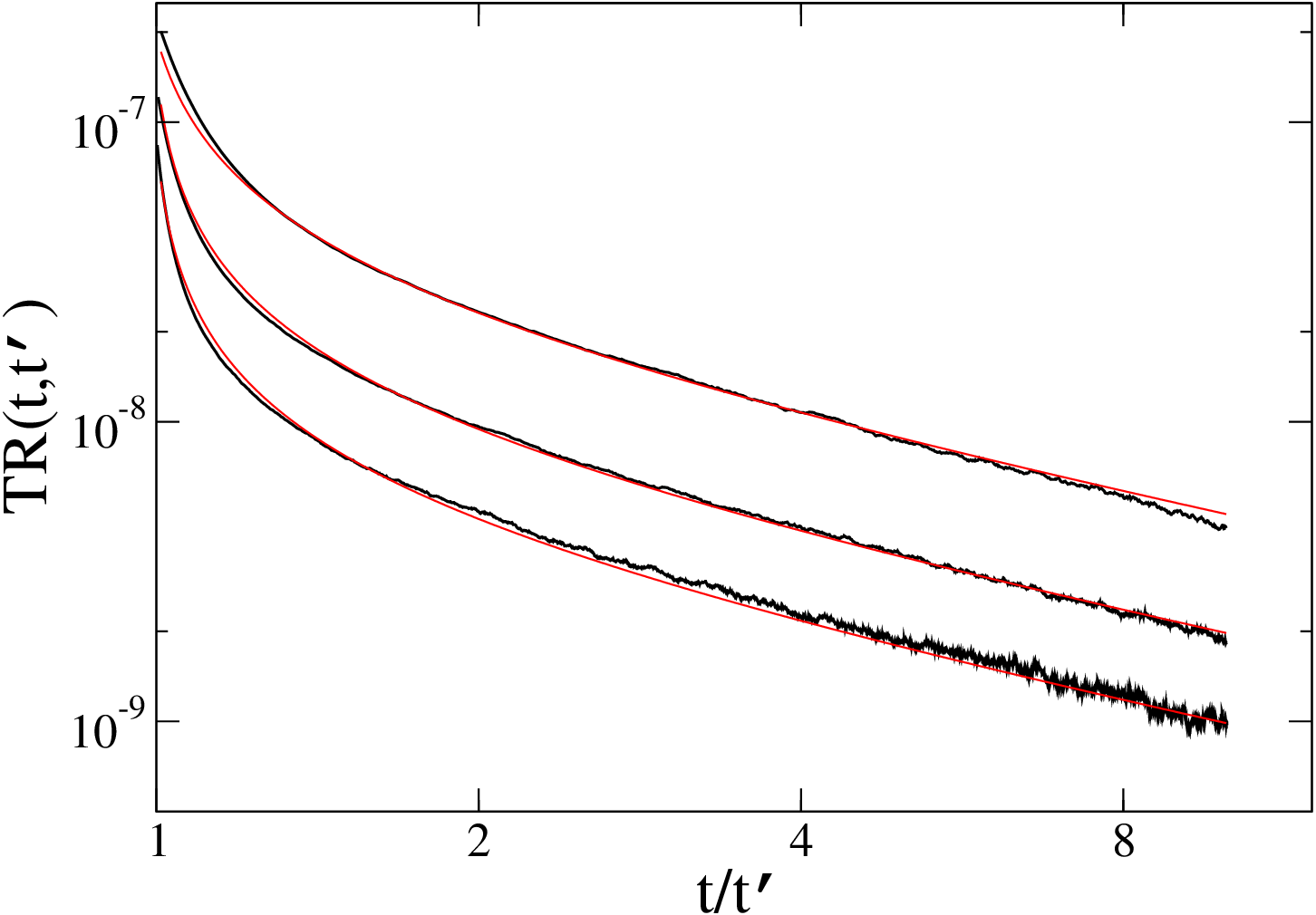}}}
    \caption{$R (t,t')$ in the $d=1$ Ising
    model with COP, $T=0.3 J$ and $J=1$. The number of spins is $N=10^4$,
    $t'=10^7,2.5 \cdot 10^7,5 \cdot 10^7$ (mcs) from top to bottom.
    Continuous curves are the plots of Eqs.~(\ref{d1.3},\ref{dd1.3}).
    Fitting parameters are $A_3=0.24, t_0=5\cdot 10^5$.}

\label{rcop}
\end{figure}

\newpage

\begin{figure}
    \centering
    \rotatebox{0}{\resizebox{.65\textwidth}{!}{\includegraphics{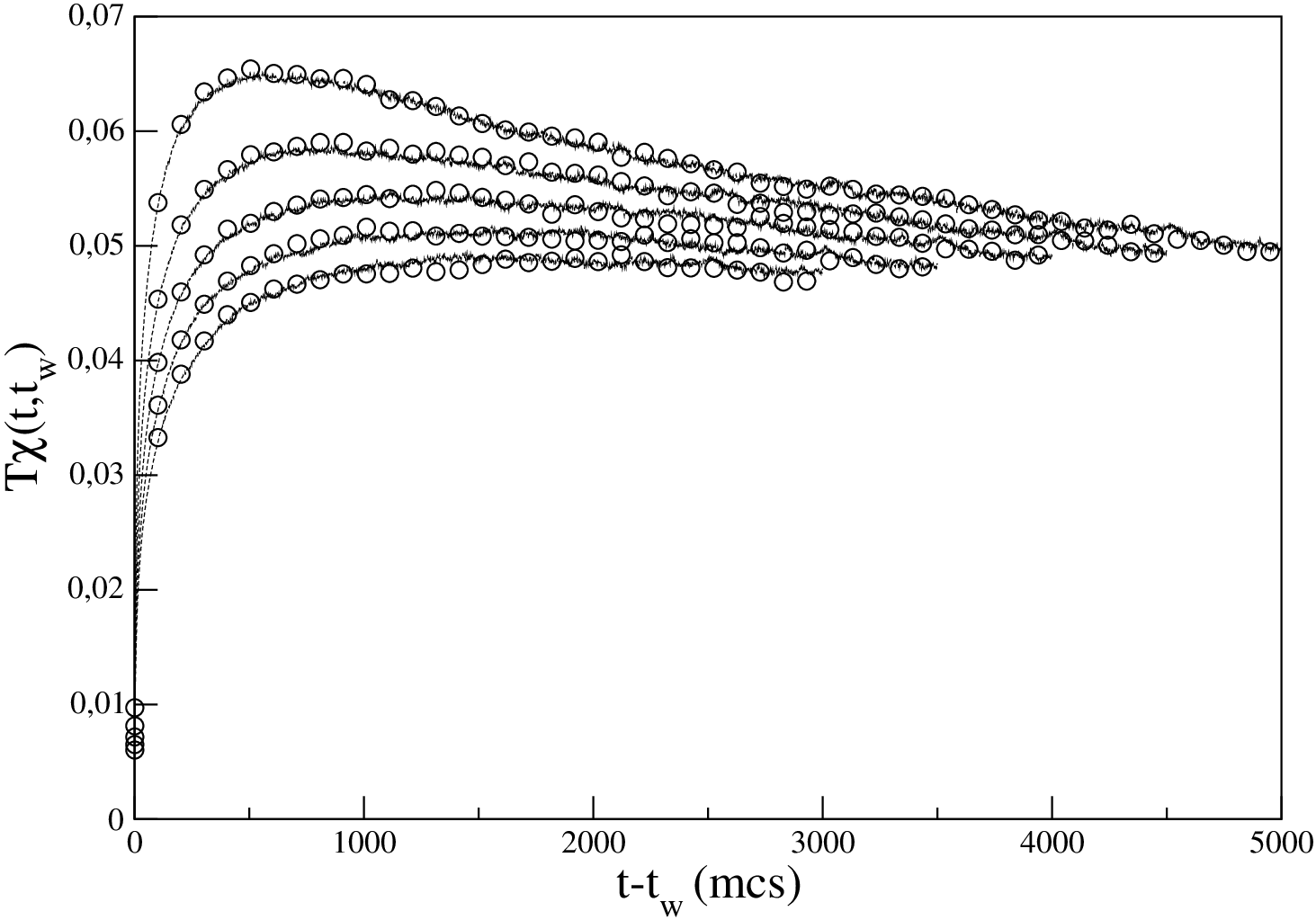}}}
    \caption{$\chi (t,t_w)$ in the $d=2$ Ising
    model with NCOP, $T=J=1$. The number of spins is $N=1600^2$,
    $t_w=1\cdot 10^3,t_w=1.5\cdot 10^3,t_w=2\cdot 10^3,t_w=2.5\cdot 10^3,t_w=3\cdot 10^3,$ 
    from top to bottom. Circles represent data obtained with
    the algorithm of Eq.~(\ref{new}), continuous lines are the results 
    with the Ricci-Tersenghi method~(\ref{old}).}

\label{figchi1}
\end{figure}

\newpage

\begin{figure}
    \centering
    \rotatebox{0}{\resizebox{.65\textwidth}{!}{\includegraphics{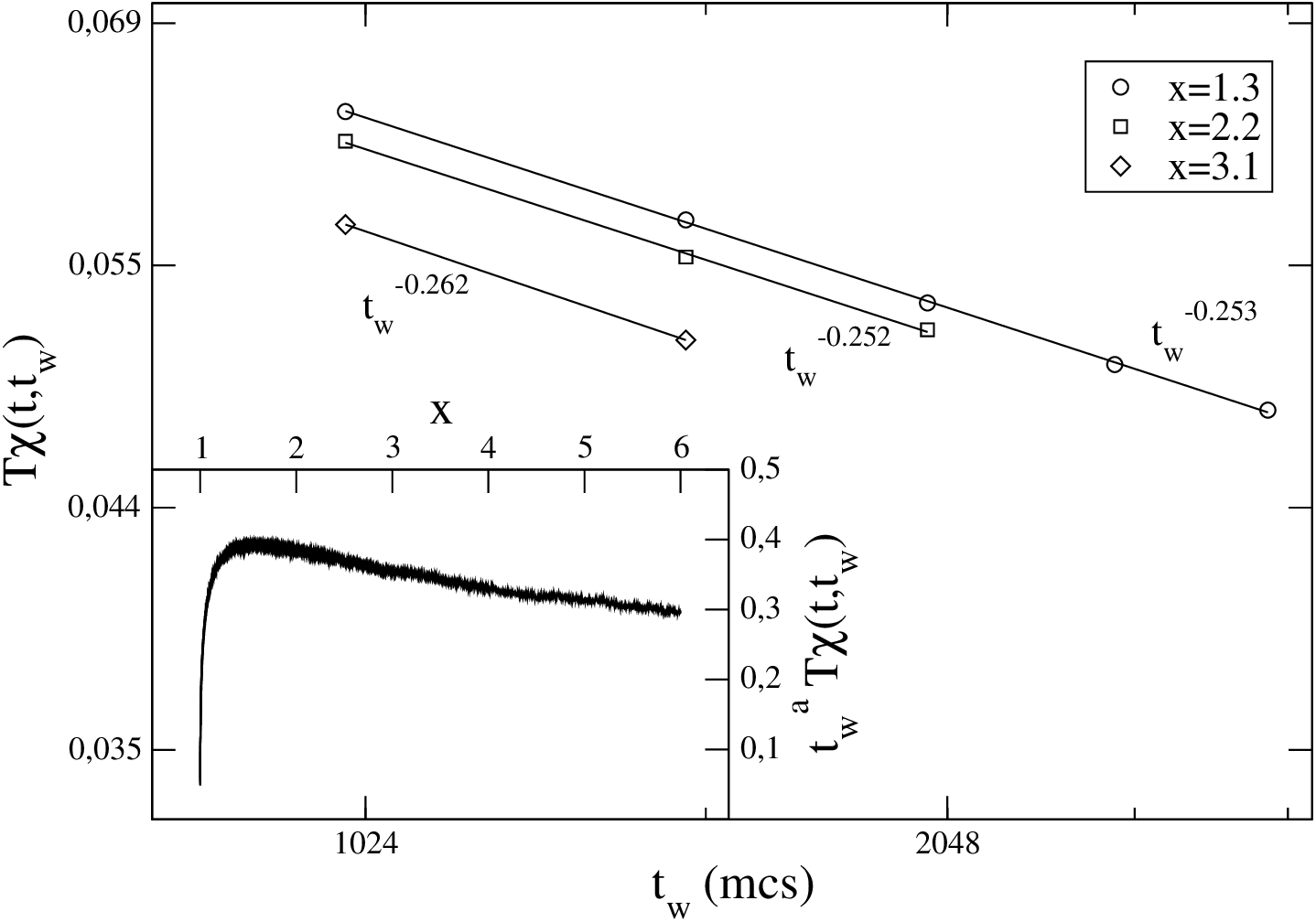}}}
    \caption{The same data of Fig.~\ref{figchi1} obtained with the algorithm of Eq.~(\ref{new}) plotted 
             for fixed values of $x$ against $t_w$. 
	     Straight lines are power law best fits.
             In the inset the data collapse for $t_w^{a}\chi(t,t_w)$, with $a= 0.26$, is shown.}

\label{figchi2}
\end{figure}


\begin{thebibliography}{000}
\bibitem{notadded}

A splitting of the response function into two parts $R^{(1)}$ and $R^{(2)}$ 
was also proposed in~\cite{crisanti}.
However, their division does not amount to a separation of a diagonal and an
off-diagonal part, as in Eqs.~(\ref{r1r1r1},\ref{erre2}). Rather,
$R^{(2)}$ is a quantity that cannot be expressed in terms of correlation functions, 
containing both diagonal and off-diagonal contributions, 
with the property that $R^{(2)}\equiv 0$ in equilibrium. $R^{(1)}$, instead,
contains only diagonal terms and can be
expressed in terms of correlation functions according to the
equilibrium FDT.
\end{thebibliography}
\end{document}